\documentclass[letterpaper, 10 pt, conference]{ieeeconf}
\IEEEoverridecommandlockouts
\usepackage{float} 
\usepackage{hyperref}
\usepackage[noend]{algpseudocode}
\usepackage{amssymb,latexsym,amsfonts,amsmath,amsthm,verbatim}
\usepackage{bbm}
\usepackage{chngcntr}
\usepackage{multicol}
\usepackage{multirow}
\usepackage{cite}
\usepackage{url,bm, mathtools, algorithm,mathrsfs,setspace,bbm,tikz,forest}
\usepackage{graphicx}
\usepackage{enumerate}

\usepackage{textcomp}
\usepackage{xcolor}
\usepackage{dsfont}

\def\BibTeX{{\rm B\kern-.05em{\sc i\kern-.025em b}\kern-.08em
    T\kern-.1667em\lower.7ex\hbox{E}\kern-.125emX}}

\newtheorem{theorem}{Theorem}
\newtheorem{lemma}{Lemma}

\newtheorem{assumption}{Assumption}

\newtheorem{example}{Example}[section]

\newcommand{\R}{{\mathbb R}}
\newcommand{\state}{\mathcal{S}}
\newcommand{\M}{\mathcal{M}}

\newcommand{\E}{{\mathbb E}}

\newcommand{\A}{\mathcal{A}}

\newcommand{\pierluigi}[1]{}
\newcommand{\RJ}[1]{}
\newcommand{\kri}[1]{}
\newcommand{\kris}[1]{}
\renewcommand{\baselinestretch}{0.94}

\begin{document}

\title{\LARGE {\bf Model-Free Reinforcement Learning for Optimal Control of Markov Decision Processes Under Signal Temporal Logic Specifications}}
\author{\normalsize Krishna C. Kalagarla, Rahul Jain,  Pierluigi Nuzzo\\
Ming Hsieh Department of Electrical and Computer Engineering, University of Southern California, Los Angeles \\
Email: \{kalagarl,rahul.jain,nuzzo\}@usc.edu
}

\maketitle

\begin{abstract}
We present a model-free reinforcement learning algorithm to find an optimal policy for a finite-horizon Markov decision process while guaranteeing a desired lower bound on the  probability of satisfying a signal temporal logic (STL) specification. We propose a method to effectively augment the MDP state space to capture the required state history and express the STL objective as a reachability objective. The planning problem can then be formulated as a finite-horizon constrained Markov decision process (CMDP). For a general finite horizon CMDP problem with unknown transition probability, we develop a reinforcement learning scheme that can leverage any model-free RL algorithm to provide an approximately optimal policy out of the general space of non-stationary randomized policies. We illustrate the effectiveness of our approach in the context of robotic motion planning for complex missions under uncertainty and performance objectives. 
\end{abstract}

\section{Introduction}

Markov decision processes (MDPs)~\cite{Puterman:1994:MDP:528623} offer a natural framework to express sequential decision-making problems and have increasingly been combined with temporal logic specifications~\cite{baier2008principles} to rigorously express complex mission objectives or constraints. In particular, signal temporal logic (STL)~\cite{maler2004monitoring} is a rich temporal extension of propositional logic that can express continuous-time continuous-valued signals and can be used, for instance, to unambiguously capture  bounds on physical variables or time-sensitive objectives.

Previous efforts have focused on maximizing the probability of satisfying a given STL specification~\cite{venkataraman2020tractable,aksaray2016q,varnai2020robustness}, for example, by maximizing a log-sum-exp approximation of the satisfaction probability.
However, in many applications, mission-critical requirements, involving stronger guarantees on the satisfaction  of  temporal logic objectives, must be paired with performance constraints, such as smoothness of motion, or fuel consumption rates, usually expressed  in  terms of cost functions. 
The focus of this paper is on these composite tasks where a total cost on an MDP must be minimized while guaranteeing a lower bound on the  probability of satisfying a given STL specification. In particular, we consider a bounded-time fragment of STL that allows up to two layers of nested temporal operators and is expressive enough to capture objectives such as ``eventually reach a location within $t_1$ minutes and remain there for $t_2$ minutes." To the best of our knowledge, this is the first paper addressing this problem formulation. 

Our contribution is twofold. We first propose a method that extends and modifies a previously proposed technique~\cite{venkataraman2020tractable} to efficiently augment the state space of the MDP and reduce STL satisfaction to a reachability objective for a finite-horizon MDP. We can then cast the logically-constrained optimal control problem as the problem of controlling a finite-horizon constrained Markov decision process (CMDP)~\cite{altman1999constrained}. As in previous approaches~\cite{aksaray2016q,venkataraman2020tractable}, we augment the MDP state space to be able to reason about the satisfaction of the STL formula. 
However, our method allows formulating probabilistic constraints on STL satisfaction and additional cost objectives which could not be expressed within a log-sum-exp formulation. 


For a general finite-horizon CMDP problem with unknown transition probability, we further introduce a model-free reinforcement learning (RL) scheme that produces an approximately optimal policy out of the general space of non-stationary randomized policies. Specifically, we formulate the CMDP problem as a min-max game between a player utilizing a no-regret algorithm and a player using a model-free RL algorithm~\cite{jin2018q,Puterman:1994:MDP:528623}. Our scheme can use any model-free RL algorithm and provides guarantees that the performance of the returned policy 
can be made arbitrarily close to that of the optimal policy. \kri{Explain better of use references. Also, what are the theoretical guarantees?}

A min-max game formulation was also used in the past to find optimal mixed deterministic policies in the context of offline RL for discounted CMDPs~\cite{le2019batch} as well as feasible policies satisfying a set of convex constraints without optimality guarantees with respect to a cost objective~\cite{miryoosefi2019reinforcement}. 
Differently from these efforts, we focus on  finite-horizon CMDPs and use the concept of occupancy measures~\cite{altman1999constrained} to obtain an approximately optimal policy out of the general space of  non-stationary randomized policies. We illustrate the applicability of our approach on two cases studies, showing that the returned policies very closely satisfy the probabilistic STL constraints and have performance comparable to that of the optimal policies. 

\kri{Wrap up with a conclusive statement on the effectiveness or evaluation of our approach. Also, I think you have to make the discussion or related approaches and contributions sharper.}

\section{Preliminaries} \label{sec:prelim}

We denote the sets of real and natural numbers by $\mathbb{R}$ and $\mathbb{N}$, respectively. $\mathbb{R}_{\geq 0}$ is the set of non-negative reals. The indicator function $\mathds{1}_{s_0}(s)$ evaluates to $1$ when $s = s_0$ and 0 otherwise. The probability simplex over the set $S$ is denoted by $\Delta_{S}$. 

\noindent \textbf{Signal Temporal Logic (STL).}
We use a fragment of signal temporal logic (STL)~\cite{maler2004monitoring}, a temporal extension of propositional logic, to specify complex tasks. The STL formulae in this paper are constructed inductively as  follows:
\begin{align}\label{eq:stldef}
    \Phi_{o} &:= \textbf{F}_{\left[0,T_{o}\right]}\Phi_{in} \ | \ \textbf{G}_{\left[0,T_{o}\right]}\Phi_{in}, \nonumber \\
    \Phi_{in} &:= \Phi_{in} \wedge \Phi_{in} \ |  \ \Phi_{in} \vee \Phi_{in} \ | \ \textbf{F}_{\left[0,T_{in}\right]}\varphi \ | \ \textbf{G}_{\left[0,T_{in}\right]}\varphi  , \nonumber \\
    \varphi &:= \mathsf{ true } \ | \ p \ | \ \neg  \varphi  \ | \ \varphi \wedge \varphi,
\end{align}  
where $T_{o}, T_{in} \in \mathbb{R}_{\geq 0}$, $\Phi_{o},\Phi_{in}$, and $\varphi$ are STL formulae, and $p$ is a predicate of the form $f(\sigma) < d $, where $ \sigma: \R_{\geq 0} \to \R^{n}$ is a signal and $f(\sigma): \R^{n} \to \R $ is a function mapping a signal to the real line. Further, $\wedge$ and $\neg$ are the logic conjunction and negation,  and $\textbf{F}$ and $\textbf{G}$ are the \emph{eventually} and \emph{always} temporal operators.

The Boolean semantics of our STL formulae are interpreted over finite length signals. Let $\sigma(t)$ be the value of the signal at time $t$, $(\sigma,t)$ be the suffix of the signal $\sigma$ starting from time $t$, and $\sigma_{t_1:t_2}$ be the segment of the signal from time $t_1$ to time $t_2$. Informally, signal $(\sigma,t)$ satisfies $p$, written $(\sigma,t) \models p$, if the predicate $p$ holds for $\sigma(t)$. The signal $(\sigma,t)$ satisfies $\textbf{F}_{\left[a,b\right]} \phi$ if there exists $a \leq t' \leq b $ such that $(\sigma,t+t')$ satisfies $\phi$. Finally, signal $(\sigma,t)$ satisfies $\textbf{G}_{\left[a,b\right]} \phi$ if $(\sigma,t+t')$ satisfies $\phi$ for all $a \leq t' \leq b $.

Let $((\sigma,t) \models \phi)$ evaluate to $1$ if true and $0$ otherwise. Then, we have the following equivalences:
\begin{align*}
    \exists \ t' \in \left[a,b\right] , (\sigma,t') \models \phi &\iff \max_{t' \in \left[a,b\right]} ((\sigma,t') \models \phi) = 1 ,\\
    \forall \ t' \in \left[a,b\right] , (\sigma,t') \models \phi &\iff \min_{t' \in \left[a,b\right]} ((\sigma,t') \models \phi) = 1,\\
    (\sigma,t) \models \phi_1 \wedge (\sigma,t) \models \phi_2 &\iff \min_{i=1,2}\{(\sigma,t) \models \phi_i \} = 1, \\
    (\sigma,t) \models \phi_1 \vee (\sigma,t) \models \phi_2 &\iff \max_{i=1,2}\{(\sigma,t) \models \phi_i\} = 1.
\end{align*}

While allowing for only two layers of nested temporal operators, this STL fragment allows specifying a rich set of time-bounded and safety requirements. The \emph{horizon} $hrz(\phi)$ \cite{dokhanchi2014line} of an STL formula $\phi$ is the minimum time length needed to certify whether a signal satisfies the formula or not. It can be computed recursively from the sub-formulae of $\phi$, as further detailed in Appendix~\ref{app:horizon}.

\noindent \textbf{Finite-Horizon MDPs.} We consider finite-horizon MDPs~\cite{Puterman:1994:MDP:528623}, which can be formally defined by a tuple $\M = (\state,\A,H,s_{0},p,c)$, where $\state$ and $\A$ denote the finite state and action spaces, respectively. The agent interacts with the environment in episodes of length $H+1$, each episode starting with the same initial state $s_{0}$. The non-stationary transition probability is denoted by $p$ where $p_{h}(s'|s,a)$ is the probability of transitioning to state $s'$ upon taking action $a$ at state $s$ at time step $h \in \{0,\ldots,H\}$. The non-stationary cost of taking action $a$ in state $s$ at time step $h $ is $c_{h}(s,a) \in \left[ 0, \bar{C} \right]$.

A non-stationary randomized policy $\pi = (\pi_{0}, \ldots , \pi_{H}) \in \Pi$, where $\pi_{i} : \state \to \Delta_{\A}$, maps each state to a probability simplex over the action space. For a state $s \in \state$ and time step $h \in \{0,\ldots,H\}$ the value function of a non-stationary randomized policy $V_{h}^{\pi}(s;c)$ is defined as $V_{h}^{\pi}(s;c) = \E \left[\sum_{i=h}^{H} c_{i}(s_{i},a_{i}) | s_h = s,\pi,p \right]$, 
where the expectation is over the environment and policy randomness. In the following, we omit $\pi$ and $c$ when they are clear from the context. The total expected cost of an episode under policy $\pi$ with respect to cost function $c$ is the respective value function from the initial state $s_0$,  i.e., $V_{0}^{\pi}(s_0;c)$. There always exists an optimal non-stationary deterministic policy $\pi^{*}$ \cite{Puterman:1994:MDP:528623} such that $V_{h}^{\pi^{*}}(s) = V_{h}^{*}(s) = \text{inf}_{\pi} V_{h}^{\pi}(s)$. 

Since the STL formulae are defined over a continuous time as opposed to discrete-step MDPs, we discretize the continuous time space by considering a step $\Delta t$. Without loss of generality, we take $\Delta t = 1$. 
A finite \emph{run} ${\xi_t}$ of the MDP at time $t \in \mathbb{N}$ is a sequence of states and actions $s_0a_0s_1,a_1\ldots s_{t}$ up to time $t$. Given an MDP $\mathcal{M}$ and an STL formula $\Phi$, a finite run $\xi_t = s_0a_0\ldots s_t$, $t \geq hrz(\Phi)$, of the MDP under policy $\pi$ is said to satisfy $\Phi$ if the
signal $s_{0:t} = s_0s_1\ldots s_{t} $ generated by the run satisfies $\Phi$. The probability that a run of $\mathcal{M}$ satisfies $\Phi$ under policy $\pi$ is denoted by $Pr^{\pi}_{\M}(\Phi)$, i.e., $ Pr^{\pi}_{\M}(\Phi) = Pr^{\pi}_{\M}(s_{0:hrz(\Phi)} \models \Phi) $.

\noindent \textbf{Finite-Horizon Constrained MDPs.} A finite-horizon constrained MDP (CMDP)~\cite{altman1999constrained} is a finite-horizon MDP with an additional constraint expressed by a pair of cost function and threshold $\{d,l \}$. For simplicity, in this paper, we consider a single constraint. Extensions to the case of multiple constraints are straightforward. 
The cost of taking action $a$ in state $s$ at time step $h \in \{0,\ldots,H\}$ with respect to the constraint cost function is $d_{h}(s,a) \in \left[ 0, \bar{D} \right]$.

Solving a CMDP problem consists in finding a policy which minimizes the total expected objective cost such that the total expected constraint cost is less than or equal to its threshold $l$. Formally,
\begin{equation}\label{eq:obj}
    \begin{aligned}
   \pi^{*} \in \underset{\pi \in \Pi }{\text{ argmin }} \quad & V_{0}^{\pi}(s_{0};c)\\ \textrm{s.t.} \quad  & V_{0}^{\pi}(s_{0};d) \leq l .
\end{aligned}
\end{equation}
The optimal value is $V^{*} = V_{0}^{\pi^{*}}(s_{0};c) $. The optimal policy may be randomized \cite{altman1999constrained}, i.e., an optimal deterministic policy may not exist as in the case of finite-horizon MDPs.

\noindent \textbf{Occupancy Measures.} Occupancy measures \cite{altman1999constrained,aaai2021} allow for an alternative representation of the set of non-stationary randomized policies and a formulation of the optimization problem \eqref{eq:obj} as a linear program (LP). The occupancy measure $q^{\pi}$ of a policy $\pi$ in a finite-horizon MDP is defined as the expected  number of visits to a state-action  pair $(s,a)$ in an episode at time step $h$. Formally, $q^{\pi}_{h}(s,a) = Pr\left[s_h= s,a_h=a|s_0 = s_0, \pi\right]$ and can be interpreted as the flow of probability through a state.

The occupancy measure $q^{\pi}$ of a policy $\pi$ satisfies linear constraints \cite{altman1999constrained} expressing non-negativity and conservation of probability flow through the states. \kri{Can you explain briefly in words what is flow conservation? You never defined it. Every time you introduce a new term, you have to clarify what you mean.} The space of the occupancy measures satisfying these constraints is denoted by $\Delta(\M)$ and is convex. A policy $\pi$ generates an occupancy measure $q \in \Delta(\M)$ if  
\begin{equation}\label{eq:occu}
\pi_h(a|s) = \frac{q_h(s,a)}{\sum_{b}q_h(s,b)}, \quad \forall (s,a,h).
\end{equation}
Thus, there exists a generating policy for all occupancy measures in $\Delta(\M)$ and \emph{vice versa}. Further, the total expected cost of an episode under policy $\pi$ with respect to cost function $c$ can be expressed in terms of the occupancy measure as $V_{0}^{\pi}(s_{0};c) = \sum_{h,s,a}q^{\pi}_{h}(s,a)c_h(s,a)$.

\section{Problem Formulation}

For a given finite-horizon MDP and STL specification, we are interested in finding a policy which minimizes the total expected cost such that the probability of satisfying the given STL specification is above a given threshold. We assume that the MDP horizon exceeds by one step the horizon of the STL specification. Our formulation can be trivially extended to longer MDP horizons. 
We then define the following problem.

\noindent \textbf{Problem 1.}
Given the MDP $\M = (\state,\A,H,s_{0},p,c)$, the STL formula $\Phi_{o}$ with horizon $H = hrz(\Phi_{o}) + 1$, and the satisfaction threshold $p_{thres}$, find a policy $\pi^{*}$ such that 
\begin{equation}\label{eq:stlobj}
    \begin{aligned}
   \pi^{*} \in \underset{\pi \in \Pi }{\text{ argmin }} \quad & \E \left[\sum_{i=0}^{H} c_{i}(s_{i},a_{i}) | s_0 = s_0,\pi \right]\\ \textrm{s.t.} \quad  
   Pr^{\pi}_{\M}(\Phi_{o}) &\geq p_{thres},
\end{aligned}
\end{equation}
where $Pr^{\pi}_{\mathcal{M}}(\Phi_{o})$ is the probability of satisfying $\Phi_{o}$ under policy $\pi$.


Because the objective in~\eqref{eq:stlobj} is not additive in nature and the dependence on the history for determining the probability of satisfying the STL formula is non-Markovian, we need to extend the state space of the MDP to capture the necessary history and evaluate the satisfaction of the formula. In the extended state space, we show that the probability of satisfaction is equal to the probability of reaching a set of states, which can be expressed by a cost function on the extended MDP. The cost function $c$ of the original MDP can also be trivially extended, leading to a standard finite-horizon CMDP formulation. We detail this reduction in Section~\ref{sec:reduction}. In Section~\ref{sec:learning}, we introduce a model-free reinforcement learning (RL) algorithm to find an $\epsilon$-optimal policy for a given finite-horizon CMDP. This algorithm is then applied to the CMDP resulting from our original problem.
\section{Reduction to CMDP}\label{sec:reduction}
The STL formula $\Phi_{o}$ is of the form $\textbf{F}_{\left[0,T_{o}\right]}\Phi_{in} $ or $\textbf{G}_{\left[0,T_{o}\right]}\Phi_{in}$. Let $\Phi_{in}$ include $n$ sub-formulae $\Phi_{in}^{i}$ of the form $\textbf{F}_{\left[0,T_{in}\right]}\varphi^{i} $ or $\textbf{G}_{\left[0,T_{in}\right]}\varphi^{i}$, $i=1,\ldots,n$. Each of these sub-formulae has horizon $hrz(\Phi_{in}^{i}) = T_{in}$, $\forall \ i,n $. Therefore, the horizon of $\Phi_{in}$ is also equal to $T_{in}$, while the one of $\Phi_{o}$ is $\tilde{H} = T_{in} + T_{o}$. We then obtain 

$s_{0:\tilde{H}} \models \Phi_{o}$
  \begin{align}\label{eq:stlsat}
       &\iff  \begin{dcases}
       \max_{t \in \left[0,T_{o}\right]} ((s,t) \models \Phi_{in}) = 1,   \Phi_{o} = \textbf{F}_{\left[0,T_{o}\right]}\Phi_{in},\\
       \min_{t \in \left[0,T_{o}\right]} ((s,t) \models \Phi_{in}) = 1,   \Phi_{o} = \textbf{G}_{\left[0,T_{o}\right]}\Phi_{in}
      \end{dcases}, \nonumber \\
       &\iff  \begin{dcases}
       \max_{ t \in \left[T_{in},\tilde{H}\right] } (s_{t-T_{in}:t} \models \Phi_{in}) = 1,    \Phi_{o} = \textbf{F}_{\left[0,T_{o}\right]}\Phi_{in}, \\
       \min_{t \in \left[T_{in},\tilde{H}\right]} (s_{t-T_{in}:t} \models \Phi_{in}) = 1,   \Phi_{o} = \textbf{G}_{\left[0,T_{o}\right]}\Phi_{in}, \nonumber \\
      \end{dcases}\\
      &\iff  \begin{dcases}
       \max_{ t \in \left[T_{in},\tilde{H}\right] } Sat(s_{t+1},\Phi_{in}),  \Phi_{o} = \textbf{F}_{\left[0,T_{o}\right]}\Phi_{in}, \\
       \min_{t \in \left[T_{in},\tilde{H}\right]} Sat(s_{t+1},\Phi_{in}),   \Phi_{o} = \textbf{G}_{\left[0,T_{o}\right]}\Phi_{in}.
      \end{dcases}
\end{align}  
\kri{this notation is confusing. iff is never defined. Can you do first one case? And the then other will come simpler. May use implications? Also, fix the formatting.}
where $Sat(s_{t},\Phi_{in})$ evaluates to $1$ if the signal segment $s_{t-T_{in}-1:t-1}$, i.e., the previous $T_{in}+1$ steps of the signal at time step $t$ satisfies $\Phi_{in}$, and evaluates to $0$ otherwise.

We introduce a flag variable $fin$ which, at time step $t+1$, is equal to $\min_{t \in \left[T_{in},\tilde{H}\right]} Sat(s_{t+1},\Phi_{in})$ for $\Phi_{o} = \textbf{G}_{\left[0,T_{o}\right]}\Phi_{in}$ and equal to $\max_{ t \in \left[T_{in},\tilde{H}\right] } Sat(s_{t+1},\Phi_{in}) $ for $\Phi_{o} = \textbf{F}_{\left[0,T_{o}\right]}\Phi_{in}$. This flag $fin$ takes values in the set $FIN =\{0,1,\bot\}$. We introduce the placeholder $\bot$ since $Sat(s_{t'},\Phi_{in})$ is undefined for $t' \leq T_{in}$. 
\kri{Sloppy paragraph. Keeps track? Can you just define the flag variable formally, with rigorous set notations, etc.?}
We similarly define $Sat(s_{t},\Phi_{in}^{i})$, $i=1,\ldots,n$, which evaluates to $1$ if the signal segment $s_{t-T_{in}-1:t-1}$ satisfies the STL formula $\Phi_{in}^{i}$ and $0$ otherwise.  By the syntax in~\eqref{eq:stldef} and the assumption that all sub-formulae have the same horizon, we can determine $Sat(s_{t},\Phi_{in})$ recursively as:
{\small
\begin{align}\label{eq:satrec}
Sat(s_{t'},\Phi_{in}^{i} \vee \Phi_{in}^{j} ) &= \max(Sat(s_{t},\Phi_{in}^{i}),Sat(s_{t},\Phi_{in}^{j})), \nonumber\\
Sat(s_{t'},\Phi_{in}^{i} \wedge \Phi_{in}^{j} ) &= \min(Sat(s_{t},\Phi_{in}^{i}),Sat(s_{t},\Phi_{in}^{j})). 
\end{align}
}
We further associate a flag $f^i$ which takes values in the set $F^{i} =\{0,1,\ldots,T_{in} \}$ with each sub-formulae $\Phi_{in}^{i}$. These flags are used to evaluate $Sat(s_{t},\Phi_{in}^{i})$ and are updated according to the following function 

$f_{t+1}^{i} =$
\begin{align}\label{eq:flagup}
\begin{dcases}
        T_{in} + 1,&\text{if } s(t) \models \varphi^{i} \text{, } \Phi_{in}^{i} = \textbf{F}_{\left[0,T_{in}\right]}\varphi^{i}, \\
       \max(f_{t}^{i} - 1 , 0),&\text{if } s(t) \not\models \varphi^{i} \text{, } \Phi_{in}^{i} = \textbf{F}_{\left[0,T_{in}\right]}\varphi^{i}, \\
      \min(f_{t}^{i} ,T_{in} ) + 1,&\text{if } s(t) \models \varphi^{i} \text{, } \Phi_{in}^{i} = \textbf{G}_{\left[0,T_{in}\right]}\varphi^{i}, \\
       0,&\text{if } s(t) \not\models \varphi^{i} \text{, } \Phi_{in}^{i} = \textbf{G}_{\left[0,T_{in}\right]}\varphi^{i}.
\end{dcases} 
\end{align}
By the definitions of $\textbf{G}$ and $\textbf{F}$, $Sat(s_{t},\Phi_{in}^{i})$ can be evaluated from $f_{t}^{i}$ as follows

$Sat(s_{t},\Phi_{in}^{i}) = $
\begin{align}\label{eq:satup}
       \begin{dcases}
        1, \quad &\text{if } f_{t}^{i} > 0 \text{, } \Phi_{in}^{i} = \textbf{F}_{\left[0,T_{in}\right]}\varphi^{i}, \quad \quad \quad \quad  \quad \quad \quad \quad \\
       0, \quad &\text{if } f_{t}^{i} = 0 \text{, } \Phi_{in}^{i} = \textbf{F}_{\left[0,T_{in}\right]}\varphi^{i}, \\
      1, \quad &\text{if } f_{t}^{i} = T_{in} + 1 \text{, } \Phi_{in}^{i} = \textbf{G}_{\left[0,T_{in}\right]}\varphi^{i}, \\
       0, \quad &\text{if } f_{t}^{i} < T_{in} + 1 \text{, } \Phi_{in}^{i} = \textbf{G}_{\left[0,T_{in}\right]}\varphi^{i}.
       \end{dcases}
\raisetag{3\normalbaselineskip}
\end{align}
By the definition of $fin$ 
we also obtain its update rule
$fin_{t+1} =$
\begin{align}\label{eq:finalflag}
    \begin{dcases}
    \bot,  & t < T_{in}, \\
    Sat(s_{t+1},\Phi_{in}), & t = T_{in}, \\
    \min( Sat(s_{t+1},\Phi_{in}), fin_{t}), &  t > T_{in} \text{, } \Phi_{o} = \textbf{G}_{\left[0,T_{o}\right]}\Phi_{in}, \\
    \max( Sat(s_{t+1},\Phi_{in}), fin_{t}), &  t > T_{in} \text{, } \Phi_{o} = \textbf{F}_{\left[0,T_{o}\right]}\Phi_{in}.
    \end{dcases}
\raisetag{4\normalbaselineskip}
\end{align}
By the definition of the flag variables above, we obtain from~\eqref{eq:stlsat} that $s_{0:\tilde{H}} \models \Phi_{o}$ if and only if $fin_{\tilde{H}+1} = 1$, 
that is, $s_{0:\tilde{H}}$ satisfies $\Phi_{o}$ if and only if the flag variable $fin$ is equal to $1$ at time $\tilde{H}+1$, where $\tilde{H} = hrz(\Phi_{o})$. The satisfaction of the specification has then been reduced to a reachability condition. 

We define a flag-augmented MDP $\M^{\times} = (\state^{\times},\A^{\times},H^{\times},s_{0}^{\times},p^{\times},d^{\times}, c^{\times})$, where $\state^{\times} = (\state \times F_1 \times \ldots \times F_n \times FIN)$, with $s^{\times} = (s,f_1,\ldots,f_n,fin)$, $\A^{\times} = \A$, $s_{0}^{\times} = (s_{0},\underset{n \text{ times }}{0,\ldots,0},\bot)$, and $H^{
\times} = hrz(\Phi_{o}) + 1$. \kris{We should revise the format here.} For the transition probability function $p^{\times}$, the $s$ component of $s^{\times}$ is updated according to the original probability transition function $p$ while the flag variables are updated according to~\eqref{eq:satrec}-\eqref{eq:finalflag}. The cost function $d^{\times}$ is defined such that the expected cost with respect to $d^{\times}$ is the probability of reaching states with flag variable $fin$ equal to 1 at time $H^{\times}$. Thus,
\begin{equation*}
    d_{h}^{\times}(s,f_1,\ldots,f_n,fin,a) = \begin{dcases}
    1, \quad \text{ if } h = H^{\times} \text{ and } fin = 1,\\
    0, \quad \text{ otherwise.}
    \end{dcases}
\end{equation*}
Similarly, the objective cost function $c$ of the original MDP can be extended to the augmented MDP $\M^{\times}$ as follows:
\begin{equation*}
    c_{h}^{\times}(s,f_1,\ldots,f_n,fin,a) = c_h(s,a). 
\end{equation*}
By the derivations above, we can state the following result. 
\begin{theorem}
For given MDP $\M=(\state,\A,H,s_{0},p,c)$, STL formula $\Phi_{o}$, and desired satisfaction threshold $p_{thres}$, Problem $1$ reduces to the following CMDP problem for the extended MDP $\M^{\times}$. 

\begin{equation}\label{eq:stlprodobj}
    \begin{aligned}
   \pi^{*} \in \underset{\pi \in \Pi^{\times} }{\text{ argmin }} \quad & V_{0}^{\pi}(s_{0}^{\times};c^{\times})\\ \textrm{s.t.} \quad  & V_{0}^{\pi}(s_{0}^{\times};d^{\times}) \geq p_{thres} .
\end{aligned}
\end{equation}
\end{theorem}

\section{The CMDP Learning Problem} \label{sec:learning}
 
We consider the setting where an agent repeatedly interacts with a finite-horizon CMDP $\M = (\state,\A,H,s_0,p,c,\{d,l\})$ in episodes of fixed length $H$, starting from the same initial state $s_0$. We assume that the cost function $c,d$ is known to the learning agent, but the transition probability $p$ is unknown. 
The main objective is to design a model-free online learning algorithm 
returning an $\epsilon$-optimal policy. A policy $\pi$ is said to be $\epsilon$-optimal if the total expected objective cost of an episode under policy $\pi$ is within $\epsilon$ of the optimal value, i.e., $V_{0}^{\pi}(s_0;c) \leq V^{*} + \epsilon$, and the constraints are satisfied within an $\epsilon$ tolerance, i.e., $V_{0}^{\pi}(s_0;d) \leq l + \epsilon$. We make the following assumption of feasibility. 

\begin{assumption}\label{asm}
The given CMDP $\M$ is feasible, i.e., there exists a policy $\pi$ such that the constraints are satisfied. 
\end{assumption}

The optimization problem \eqref{eq:obj} can be formulated in terms of occupancy measures as:

\begin{equation}\label{eq:q_obj}
    \begin{aligned}
   q^{*} \in \underset{q \in \Delta(\M) }{\text{ argmin }} \quad & C(q)\\ \textrm{s.t.} \quad  & D(q) \leq l,
\end{aligned}
\end{equation}
where $C(q) = \sum_{h,s,a}q_{h}(s,a)c_h(s,a)$ and $D(q) = \sum_{h,s,a}q_{h}(s,a)d_h(s,a)$.

The Lagrangian of this optimization problem is $L(q,\lambda) = C(q) + \lambda (D(q) - l)$, where $\lambda \in \R_{+}$ is the Lagrangian multiplier. Following  standard results from optimization theory~\cite{boyd2004convex}, the optimization problem \eqref{eq:q_obj} can be formulated as the following min-max problem: $\min_{q\in \Delta(\M)} \max_{\lambda \in \R_{+}} L(q,\lambda)$. Further, the functions $C(q)$ and $D(q)$ are linear in $q$ and the set of occupancy measures $\Delta(\M)$ expressed by linear constraints is convex. Therefore, by strong duality~\cite{boyd2004convex}, the optimization problem \eqref{eq:obj} is also equivalent to the max-min problem $ \max_{\lambda \in \R_{+}} \min_{q\in \Delta(\M)} L(q,\lambda)$.

The latter problem can be viewed as a zero-sum game between a $\lambda$-player, who seeks to maximize $L(q,\lambda)$, and a $q$-player, who seeks to minimize $L(q,\lambda)$. We use a previously proposed approach~\cite{freund1999adaptive} for solving such a game. In this approach, the $\lambda$-player plays a no-regret online learning algorithm \cite{hazan2019introduction} against the \textit{best response} strategy played by the $q$-player. In no-regret online learning, the difference between  the  cumulative gain of the player and that of the best fixed decision in hindsight is sub-linear in the number of plays or iterations. Specifically, for each $t$, given $\lambda_t$ played by the $\lambda$-player, the $q$-player plays the \textit{best response} $q_t$ with respect to the loss function $L(q,\lambda_t)$. The $\lambda$-player then observes the gain function $l_{t}(\lambda)$, which is the Lagrangian $L(q_t,\lambda) = C(q_t) + \lambda (D(q_t) - l)$. With this feedback, the  $\lambda$-player updates the Lagrange multiplier $\lambda$ according to a no-regret online learning algorithm. We refer to Appendix~\ref{app:ol} for further details.

The \textit{best response} above is the occupancy measure which minimizes the current Lagrangian $L(q,\lambda_t)$, i.e.,
\begin{align*}
 &\underset{q \in \Delta(\M) }{\text{ argmin }} L(q,\lambda_t) = 
    \underset{q \in \Delta(\M) }{\text{ argmin }}C(q) + \lambda_{t}(D(q)- l).
\end{align*}
This best response can be calculated by finding the optimal policy of the MDP with respect to cost function $ c + \lambda_{t}d$. The optimal policy is then translated into its associated occupancy measure which is the desired best response. 
\kri{The text above is a bit dense. Some steps may need rephrasing or additional clarifications.}

\subsection{Occupancy Based Model-Free Constrained Reinforcement Learning ({\tt OB-MFC}) Algorithm}

\begin{algorithm}[t]
\begin{algorithmic}[0]
\caption{{\tt Meta-Algorithm}}
\label{alg:meta}
\State Initialize $\lambda_1$
\For {$t=1,\ldots,T$}
\State $q_{t} \leftarrow \texttt{Best-Response}(\lambda_t)$,
\State $\lambda$-player is given the gain function $L(q_t,\lambda)$,
\State $\lambda_{t+1} \leftarrow \texttt{OnlineLearning}(\lambda_{1},q_{1},\ldots,\lambda_{t},q_{t})$.
\EndFor 
\State \textbf{Return} $\frac{1}{T}\sum_{t=1}^{t=T}q_t$. \end{algorithmic}
\end{algorithm}

We summarize the above approach in Algorithm~\ref{alg:meta}.
The \texttt{Best-Response} function can be implemented by using any model-free RL algorithm~\cite{jin2018q,Puterman:1994:MDP:528623} to find an optimal policy with respect to a scalar cost function $c + \lambda_{t}d$. To ensure finite completion time for the RL algorithm, we can make the simple assumption that the RL algorithm \texttt{Best-Response-Policy} returns an $\epsilon$-optimal policy. 

\begin{assumption}\label{asm:br}
Given cost functions $c,d$ and $\lambda \in \R_{+}$, the RL algorithm \texttt{Best-Response-Policy} returns a policy $\pi$ such that $V(\pi) < \min_{\pi^{'} \in \Pi} V(\pi^{'}) + \epsilon_{br}$, where $V(\pi)$ is the total expected return with respect to cost function $c + \lambda d$.
\end{assumption}

The corresponding occupancy measure $q_t$ of policy $\pi_t$ can be estimated by Monte Carlo estimate following the definition of an occupancy measure, i.e., $q^{\pi}_{h}(s,a) = Pr\left[s_h= s,a_h=a|s_0 = s_0, \pi\right]$. We make a further assumption that an \texttt{Occupancy-Estimator} returns a good estimate of the occupancy measure. 

\begin{assumption}\label{asm:est}
Given policy $\pi$, \texttt{Occupancy-Estimator} returns an occupancy measure estimate $\hat{q}$  such that $\|q - \hat{q}  \|_{1} \leq \epsilon_{oe}$, where $q$ is the occupancy measure of policy $\pi$.
\end{assumption}

Most online convex optimization algorithms \cite{hazan2019introduction} make a decision from a bounded convex space. We thus require $ \lambda \leq B$, where $B$ is a hyper-parameter to be chosen. The scalar $\lambda$ is then augmented by one more dimension corresponding to $B - \lambda$ to give a bidimensional vector $(\lambda[1],\lambda[2])$. The cost function $d$ can also be seen as being augmented by $0$. The online learning agent then chooses $\lambda$ such that $||\lambda||_1 = B$.

We use the Exponentiated Gradient (EG)~\cite{kivinen1997exponentiated} online learning algorithm, which is known to be a no-regret algorithm. This algorithm utilizes the sub-gradient $\partial l_t $ of the revealed gain function $l_t(\lambda)$, namely, $ L(\hat{q},\lambda) = C(\hat{q}_t) + \lambda[1] (D(\hat{q}_t) - l)$,  which is nothing but $ \left[  (D(\hat{q}_t)-l) , 0 \right]^{T} $. We denote by $\hat{q}_t$ the estimate of $q_t$, the occupancy measure associated with $\pi_t$, obtained by \texttt{Occupancy-Estimator}. This estimate is used to approximate the sub-gradient $\partial l_t $ by using $ D(\hat{q}_t) = \sum_{h,s,a} \hat{q}_h(s,a) d_h(s,a)$. 
By putting all this together, \kri{a bit hard to follow though. I suggest you should read again after you write a section.} we obtain the occupancy-based model-free constrained reinforcement learning ({\tt OB-MFC}) Algorithm~\ref{alg:obmfc}.
\begin{algorithm}[t]
\begin{algorithmic}[0]
\caption{{\tt OB-MFC Reinforcement Learning}}
\label{alg:obmfc}
\State \textbf{Input:} Bound $B$, learning rate $\eta$, number of roll-outs $N$, number of iterations $T$.
\State Initialize $\lambda = (\frac{B}{2},\frac{B}{2}).$
\For {$t=1,\ldots,T$}
\State $\pi_{t} \leftarrow \texttt{Best-Response-Policy}(\lambda_t)$,
\State $\hat{q}_t \leftarrow \texttt{Occupancy-Estimator}(\pi,N) $,
\State $D(\hat{q}_t) \leftarrow \sum_{h,s,a} \hat{q}_h(s,a) d_h(s,a)$,
\State $\beta_t = \left[  \left(D(\hat{q}_t)-l \right) , 0 \right]^{T} $,
\State $\lambda_{t+1}\left[ i \right] \leftarrow B \frac{\lambda_{t}\left[ i \right]e^{\eta\beta_t\left[ i \right]}}{\sum_{j}\lambda_{t}\left[ j \right]e^{\eta\beta_t\left[ j \right]}} \quad $ for $i=1,2.$ 
\EndFor
\State $\tilde{q} \leftarrow \frac{1}{T}\sum_{t=1}^{T} \hat{q}_t $,
\State $\tilde{\pi}_h(a|s) \leftarrow \frac{\tilde{q}_h(s,a)}{\sum_{b}\tilde{q}_h(s,b)}, \quad \forall (s,a,h)$.
\State \textbf{Return} $\tilde{\pi}$.
\end{algorithmic}
\end{algorithm}
\subsection{Optimality of OB-MFC RL Algorithm}

In this section, we provide guarantees that the performance of the returned policy with respect to the given CMDP problem can be arbitrarily close to that of the optimal policy. The proofs of the following results can be found in the appendix.

We first show that the difference between the Lagrangian functions $L(q,\lambda)$ and $L(\hat{q},\lambda)$ with respect to the true occupancy measure $q$ and the  estimated occupancy measure $\hat{q}$ is small.

\begin{lemma}\label{lem:qerror}
Let $q$ be the occupancy measure associated with a policy $\pi$ and $\hat{q}$ be its empirical estimate such that the $L_1$ estimation error is small, i.e., $\|q - \hat{q}  \|_{1} \leq \epsilon_{oe}$. Then,  $|L(\hat{q},\lambda) - L(q,\lambda)| \leq \epsilon_{est}$ for all $||\lambda||_1 = B $, where   $\epsilon_{est} = (\bar{C} + B\bar{D})\epsilon_{oe}$.
\end{lemma}
We next show that the primal-dual gap falls below a desired threshold $\epsilon_{ol}$ after a suitably large number of iterations $T$ of the algorithm.

\begin{lemma}\label{lem:primdualgap}
After $T$ iterations of the algorithm, we have 
\begin{align*}
    \underset{\lambda \in R_{+}^{2}, ||\lambda||_1 = B}{\max} L(\tilde{q},\lambda) - L(\bar{q},\tilde{\lambda}) & \leq \epsilon_{ol}/2, \\
    \underset{q \in \Delta(\M)}{\min} L(q,\tilde{\lambda})- L(\bar{q},\tilde{\lambda}) &\geq \epsilon_{ol}/2.
\end{align*}
  Thus, the following holds for the primal-dual gap 
  $$\underset{\lambda \in R_{+}^{2}, ||\lambda||_1 = B}{\max} L(\tilde{q},\lambda) - \underset{q \in \Delta(\M)}{\min} L(q,\tilde{\lambda})  \leq  \epsilon_{ol},$$ where $\tilde{q} = \frac{1}{T}\sum_{t=1}^{T} \hat{q}_t $, $\bar{q} = \frac{1}{T}\sum_{t=1}^{T} {q}_t $, $\tilde{\lambda} = \frac{1}{T}\sum_{t=1}^{T} \lambda_t,$ and $\epsilon_{ol} = 2\epsilon_{br} + 2\epsilon_{est} + \frac{o(T)}{T}$.
\end{lemma}
We can thus make the primal dual gap arbitrarily small by reducing $\epsilon_{br}$ and $\epsilon_{est}$ and increasing the number of iterations $T$. Let the number of iterations $T$ be large enough such that $\frac{o(T)}{T} < \epsilon_{reg}$. Then, we obtain $\epsilon_{ol} < \epsilon_{reg} + 2\epsilon_{br} + \epsilon_{est}$. We now show that the returned average occupancy measure approximately satisfies the constraint and has an expected return close to that of the optimal policy.

\begin{lemma}\label{lem:error}
Under Assumption \ref{asm}, the returned occupancy measure estimate $\tilde{q}$ approximately satisfies the given constraint
$$ D(\tilde{q}) \leq l + \frac{ 2(\bar{C}(H+1) + \epsilon_{ol} + \epsilon_{est})}{B}. $$
Further, the objective value returned by $\tilde{q}$ is close to that of the optimal policy
$$ C(\tilde{q}) \leq C(q^{*}) + \epsilon_{reg} + \epsilon_{br} +\epsilon_{est}. $$
\end{lemma}

The returned $\tilde{q}$ is an estimate of the desired occupancy measure $\bar{q}$. Thus, $\tilde{q}$ may not be a valid occupancy measure, i.e., it may not correspond to a valid policy. Nevertheless, we show that the occupancy measure associated with the policy $\tilde{\pi}$ generated from $\tilde{q}$ is close to $\bar{q}$. 

\begin{lemma}\label{lem:occuerror}
Let $\bar{q}$ be the occupancy measure associated with a policy $\pi$ and $\tilde{q}$ be its empirical estimate such that the $L_1$ estimation error is small, i.e., $\|\bar{q} - \tilde{q}  \|_{1} \leq \epsilon_{oe}$. Then, for a policy defined as $\tilde{\pi}_h(a|s) = \frac{\tilde{q}_h(s,a)}{\sum_{b}\tilde{q}_h(s,b)}, \forall (s,a,h)$, the $L_1$ error between its associated occupancy measure $\tilde{\tilde{q}}$ and $\bar{q}$ is also small, i.e., $\|\tilde{\tilde{q}} - \bar{q}  \|_{1} \leq 2(H+1) \epsilon_{oe}$. Furthermore, $\|\tilde{\tilde{q}} - \tilde{q}  \|_{1} \leq (2H+3) \epsilon_{oe}$ holds. 
\end{lemma}

From Lemmas \ref{lem:error} and  \ref{lem:occuerror}, we have:

\begin{theorem}\label{thm:final}
Under Assumption \ref{asm}, the returned policy $\tilde{\pi}$ approximately satisfies the given constraint
$$ D(\tilde{\pi}) \leq l + \bar{D}(2H+1)\epsilon_{oe}+ \frac{ 2(\bar{C}(H+1) + \epsilon_{ol} + \epsilon_{est})}{B}.  $$
Further, the expected objective cost under $\tilde{\pi}$ is close to that of the optimal policy, i.e., 
$$ C(\tilde{\pi}) \leq C(\pi^*) + \bar{C}(2H+3)\epsilon_{oe}+ \epsilon_{reg} + \epsilon_{br} +\epsilon_{est}.$$
\end{theorem}

The above result shows that the performance of the returned policy can be made arbitrarily close to that of the optimal policy by making the errors $\epsilon_{ol}, \epsilon_{est}, \epsilon_{reg}$ arbitrarily small and the Lagrange multiplier bound $B$ suitably large. We can attain arbitrarily small errors $\epsilon_{ol}, \epsilon_{est}$ and  $\epsilon_{reg}$ by using a sufficiently large number of iterations $T$, a better \texttt{Occupancy-Estimator} (i.e., a larger number of roll-outs for better Monte Carlo estimation), and by running the model-free RL algorithm longer to obtain a policy closer to the optimal best response.

\section{Experimental Results}

We implemented our framework in \textsc{Python} and used the LP solver provided by \textsc{Gurobi} to find the optimal cost of a CMDP with a known transition probability. We evaluate our framework on two case studies involving motion planning of a mobile robot. The experiments are run on a $1.4$-GHz Core i5 processor with $16$-GB memory.

\begin{figure}
    \centering
    \includegraphics[width=0.35\linewidth]{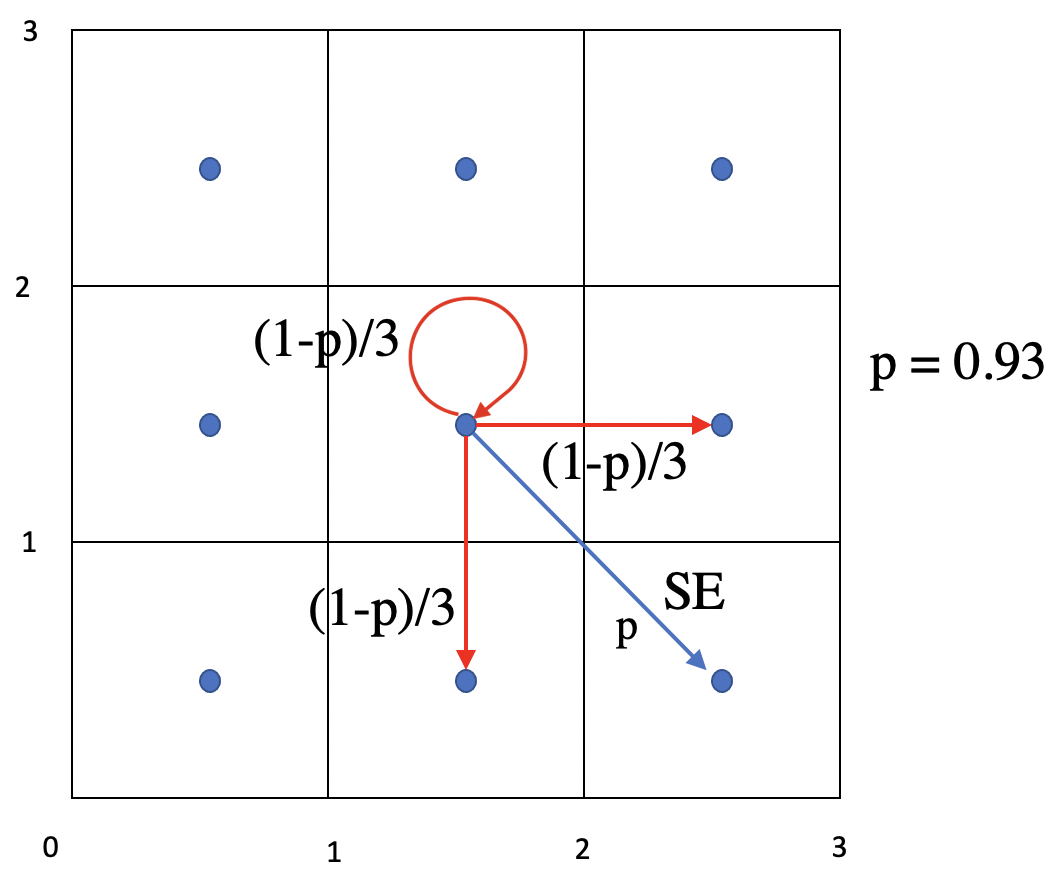}
    \caption{Robot dynamics for action $\mathsf{SE}$.}
    \vspace{-15pt}
    \label{fig:grid}
\end{figure}

We consider a robot moving with discrete actions in a simple grid world with discrete states as shown in Fig.~\ref{fig:grid}. The set of actions available to the robot in each state is $\A = (\mathsf{N},\mathsf{E},\mathsf{S},\mathsf{W},\mathsf{NE},\mathsf{NW},\mathsf{SE},\mathsf{SW},\mathsf{rest})$. The dynamics of the robots is as follows. The action $\mathsf{rest}$ does not change the robot state. Also, if the robot cannot move in the intended direction, then it remains in the same state. For all other actions, the robot moves in the intended direction with probability $p = 0.93$ and the remaining probability is equally divided between the following choices: the two possible adjoining directions and staying in the same state, as shown in Fig.~\ref{fig:grid}. 
For all time steps and states, the cost of action $\mathsf{rest}$ is $0$, the cost of the horizontal or vertical actions, i.e., $(\mathsf{N},\mathsf{E},\mathsf{S},\mathsf{W})$ is $1$, and the cost of the diagonal actions,  i.e., $\mathsf{NE},\mathsf{NW},\mathsf{SE},\mathsf{SW}$ is $2$.
We use the standard model-free Q-learning~\cite{Puterman:1994:MDP:528623} algorithm to implement \texttt{Best-Response-Policy} and Monte-Carlo estimation with $5,000$ trajectories to implement \texttt{Occupancy-Estimator}.

\subsection{Case Study 1: Bounded-Time Reachability}

In this case study, we consider a grid world of size $(6 \times 6)$ with the robot starting at $(0.5,0.5)$. The STL formula $\Phi_{o} = \textbf{F}_{\left[0,7\right]}\textbf{G}_{\left[0,1\right]} (x > 4 \wedge y > 4)$ expresses a requirement of the form ``Eventually visit and remain for $t_1$ units of time in the desired region within $t_2$ units of time." The horizon of the MDP problem is $hrz(\Phi_{o}) + 1 = 9$.

We construct the extended MDP as described in  Section~\ref{sec:reduction}, resulting in an extended state space $\state^{\times}$ with $|\state^{\times}| = 324$, and consider two different thresholds for STL satisfaction $p_{thres}$, i.e., $0.5$ and $0.9$. Since the transition probability is known by construction in both these cases, an optimal policy and the true optimal total cost is obtained by solving the LP formulation of the finite horizon CMDP as described in Section~\ref{sec:reduction}. The optimal cost for $p_{thres}=0.5$ and $p_{thres} = 0.9$ is $5.881$ and $7.494$, respectively.

In the more difficult setting of unknown transition probability, an optimal policy is obtained by using the model-free \texttt{OB-MFC} algorithm. The resulting policies are used to generate $10,000$ trajectories, and the satisfaction probabilities and expected total costs are estimated. 
The estimated satisfaction probability for $p_{thres} = 0.5$ and $p_{thres} = 0.9$ is $0.501$ and $0.897$, respectively. The estimated total expected cost for $p_{thres} = 0.5$ and $p_{thres}=0.9$ is $6.284$ and $7.589$, respectively. 

In both cases, the estimated satisfaction probability and total expected cost of the returned policy is within a small, $6.8\%$ tolerance from the optimal value and satisfaction threshold.

\subsection{Case Study 2: Bounded Time Patrolling}

In this case study, we consider a grid world of size $(4 \times 4)$ with the robot starting at $(1.5,1.5)$. The STL formula $\Phi_{o} = \textbf{G}_{\left[0,12\right]}\big(\textbf{F}_{\left[0,2\right]} (x>1 \wedge x<2 \wedge y>3 \wedge y<4) \wedge \textbf{F}_{\left[0,2\right]} (x>2 \wedge x<3 \wedge y>2 \wedge y<3) \big)$ expresses a requirement of the form  ``For all time $t \in \left[0,t_1\right]$, eventually visit region A in interval $\left[t,t +h \right]$ and eventually visit region B in interval $\left[t,t +h \right]$." The horizon of the MDP problem is $hrz(\Phi_{o}) + 1 = 15$.

Similarly to the first case study, we construct an extended state space $\state^{\times}$ with $|\state^{\times}| = 768$ and consider a threshold for STL satisfaction $p_{thres} = 0.7$. For known transition probability, an optimal policy and the true optimal total cost are obtained by solving the LP formulation of the finite-horizon CMDP as described in Section~\ref{sec:reduction}. The optimal cost for $p_{thres}=0.7$ is $16.875$.

In the more difficult setting of unknown transition probability, an optimal policy is obtained by using the \texttt{OB-MFC} algorithm. The resulting policy is used to generate $10,000$ trajectories and the satisfaction probability and expected total cost are estimated. The estimated satisfaction probability for $p_{thres} = 0.7$ is $0.702$ and the estimated total expected cost is $17.215$. The estimated satisfaction probability of the returned policy satisfies the given threshold and the estimated total expected cost is within a small, $2.01\%$ tolerance from the optimal value. 

\section{Conclusions}

We designed and validated a model-free reinforcement learning algorithm for a general finite-horizon constrained Markov decision process (CMDP) and applied it to find a cost-optimal policy for a finite-horizon Markov decision process such that the probability of satisfying a given signal temporal logic specification is beyond a desired threshold. Future plans include the extension of the proposed method to more general STL specifications and the optimization of the robust satisfaction of STL formulae. 

\bibliographystyle{IEEEtran}
\bibliography{ref}

\newpage

\appendix

\subsection{Horizon of STL Formulae}\label{app:horizon}

The \emph{horizon} $hrz(\phi)$ \cite{dokhanchi2014line} of an STL formula $\phi$ is the minimum time length needed to certify whether a signal satisfies $\phi$ or not. It is computed recursively from the sub-formulae of $\phi$ as follows:
\begin{align*}
    hrz(p) &= 0, \\
    hrz(\phi_1 \wedge \phi_2) &= max\{hrz(\phi_1),hrz(\phi_2)\},\\
    hrz(\phi_1 \vee \phi_2) &= max\{hrz(\phi_1),hrz(\phi_2)\},\\
    hrz(\textbf{F}_{\left[a,b\right]}\phi) &= b + hrz(\phi), \\
    hrz(\textbf{G}_{\left[a,b\right]}\phi) &= b + hrz(\phi), 
\end{align*}
where $p$ is a predicate and $a,b \in \mathbb{R}_{\geq 0}$.

\subsection{Online Learning}\label{app:ol}

In the framework of online convex optimization, for $t=1,\ldots,T$, an agent plays decision $\lambda_t$ belonging to a convex set $\Lambda$, following which the environment reveals a gain function $l_t: \Lambda \to \R$ such that the agent gains $l_t(\lambda_t)$. The agent attempts to minimize the regret $R_T$ which is defined as the difference between the cumulative gain of the agent and that of the best fixed decision in hindsight, i.e., 
$$ R_T = \max_{\lambda \in \Lambda} \left[ \sum_{t=1}^{T}  l_t(\lambda) \right] -  \left[ \sum_{t=1}^{T}  l_t(\lambda_t) \right].$$

\noindent An algorithm is said to be no-regret if its regret $R_T$ is $o(T)$, i.e., sub-linear in $T$.

\subsection{Proof of Lemma \ref{lem:qerror}}
We obtain
\begin{align*}
    |L(\hat{q},\lambda) - L(q,\lambda)| &= |C(\hat{q}) + \lambda[1] D(\hat{q}) - C({q}) - \lambda[1] D({q})|\\
    &= |C(\hat{q}) - C(q) + \lambda[1](D(\hat{q}) - D(q))|\\
    &\leq |C(\hat{q}) - C(q)| + \lambda[1] |D(\hat{q}) - D(q)|.
\end{align*}
The cost functions $c$, $d$ are bounded above by $\bar{C}$ and $\bar{D}$, respectively. $||\lambda||_1 = B$ and $\|q - \hat{q}  \|_{1} \leq \epsilon_{oe}$. Thus, by Cauchy-Schwartz inequality, we have 
$|L(\hat{q},\lambda) - L(q,\lambda)| \leq (\bar{C} + B\bar{D})\epsilon_{oe}$ for all $||\lambda||_1 = B $.

\subsection{Proof of Lemma \ref{lem:primdualgap}}

We have $\underset{\lambda \in R_{+}^{2},||\lambda||_1 = B}{\max} L(\tilde{q},\lambda) $
\allowdisplaybreaks\begin{align*}
    & = \frac{1}{T}\underset{\lambda \in R_{+}^{2},||\lambda||_1 = B}{\max} \sum_{t} L(\hat{q}_t,\lambda) \text{ (by linearity of }L)\\
    &\leq \frac{1}{T} \left[ \sum_{t} L(\hat{q}_t,\lambda_t) + o(T) \right] \\ &\text{(by the no-regret property of the EG algorithm})\\
    &\leq \frac{1}{T} \sum_{t} L(q_t,\lambda_t) + \epsilon_{est} + \frac{o(T)}{T}\quad \text{(by Lemma \ref{lem:qerror})}\\
    &\leq \frac{1}{T} \sum_{t} \left[ \underset{q\in \Delta(\M)}{\min} L(q,\lambda_t) \right] + \epsilon_{br} + \epsilon_{est} + \frac{o(T)}{T}\quad \\
    & \text{(by Assumption \ref{asm:br})}\\
    &\leq \frac{1}{T} \sum_{t}  L(\bar{q},\lambda_t) + \epsilon_{br} + \epsilon_{est} + \frac{o(T)}{T} \quad (\text{where }\bar{q} = \frac{\sum_{t}q_t}{T})\\
    &= L(\bar{q},\tilde{\lambda}) + \epsilon_{br} + \epsilon_{est} + \frac{o(T)}{T} \quad \text{(by linearity of $L$).} 
    \end{align*}

Similarly, we have,
\allowdisplaybreaks\begin{align*}
&\underset{q \in \Delta(\M)}{\min} L(q,\tilde{\lambda}) \\&= \frac{1}{T} \underset{q \in \Delta(\M)}{\min} \sum_{t} L(q,\lambda_t) \quad \text{(by linearity of $L$)}\\
&\geq \frac{1}{T} \sum_{t} \underset{q \in \Delta(\M)}{\min} L(q,\lambda_t)\\
&= \frac{1}{T} \sum_{t} L(q_t,\lambda_t) - \epsilon_{br} \quad \text{(by Assumption \ref{asm:br})}\\
&\geq \frac{1}{T} \sum_{t} L(\hat{q}_t,\lambda_t) - \epsilon_{br} - \epsilon_{est} \quad \text{(by Lemma \ref{lem:qerror})}\\
&\geq \frac{1}{T} \underset{\lambda \in R_{+}^{2},||\lambda||_1 = B}{\max} \sum_{t} L(\hat{q}_t,\lambda) - \frac{o(T)}{T}  - \epsilon_{br} - \epsilon_{est} \\ &\text{(by the no-regret property of the EG algorithm})\\
&\geq \frac{1}{T}\sum_{t} L(\hat{q}_t,\tilde{\lambda}) - \frac{o(T)}{T}  - \epsilon_{br} - \epsilon_{est}\\
&= L\left(\frac{\sum_{t} \hat{q}_t}{T}  ,\tilde{\lambda}\right) - \frac{o(T)}{T} - \epsilon_{br} - \epsilon_{est} 
\quad \text{(by linearity of $L$)}\\
&= L(\bar{q},\tilde{\lambda}) - \frac{o(T)}{T} - \epsilon_{br} - \epsilon_{est}
\end{align*}

Putting together the results above, we have $$\underset{\lambda \in R_{+}^{2}, ||\lambda||_1 = B}{\max} L(\tilde{q},\lambda) - \underset{q \in \Delta(\M)}{\min} L(q,\tilde{\lambda})  \leq  2\epsilon_{br} + 2\epsilon_{est} + \frac{o(T)}{T}.$$

\subsection{Proof of Lemma \ref{lem:error}}

By lemma \ref{lem:primdualgap}, we have $\underset{\lambda \in R_{+}^{2}, ||\lambda||_1 = B}{\max} L(\tilde{q},\lambda) - L(\bar{q},\tilde{\lambda}) \leq \epsilon_{ol}/2$
\allowdisplaybreaks\begin{align*}
    \implies & \underset{\lambda \in R_{+}^{2}, ||\lambda||_1 = B}{\max} L(\tilde{q},\lambda) \leq L(\bar{q},\tilde{\lambda}) + \epsilon_{ol}/2 \\
    \implies & C(\tilde{q}) + \underset{\lambda \in R_{+}^{2}, ||\lambda||_1 = B}{\max}  \lambda[1](D(\tilde{q})-l) \\
    & \leq C(\bar{q}) + \tilde{\lambda}[1] (D(\bar{q})-l)  +  \epsilon_{ol}/2.
\end{align*}
Since $|q_t - \hat{q}_t| \leq \epsilon_{oe}$, we have $|\bar{q} - \tilde{q}| \leq \epsilon_{oe}$ by triangular inequality. Thus, $C(\bar{q}) \leq C(\tilde{q}) + \bar{C}\epsilon_{oe}$ holds by Cauchy-Schwartz inequality. Finally, we obtain 
$$ \underset{\lambda \in R_{+}^{2}, ||\lambda||_1 = B}{\max}  \lambda[1](D(\tilde{q})-l) \leq \tilde{\lambda}[1] (D(\bar{q})-l) +  \epsilon_{ol}/2 + \bar{C}\epsilon_{oe}.$$
If $D(\tilde{q}) \leq l$ holds, we have trivially shown that the constraint is satisfied. Else, by taking the maximum possible value of $\lambda[1]$, we have
\begin{equation}\label{eq:lem3_1}
  B(D(\tilde{q})-l) \leq \tilde{\lambda}[1] (D(\bar{q})-l) +  \epsilon_{ol}/2 + \bar{C}\epsilon_{oe}.   
\end{equation}

By Assumption~\ref{asm}, there exists $q^{f}$ such that $D(q^{f}) \leq l$. Then, we have $L(\bar{q},\tilde{\lambda}) -  L(q^{f},\tilde{\lambda})$
\begin{align*}
    &\leq L(\tilde{q},\tilde{\lambda}) -  L(q^{f},\tilde{\lambda}) + \epsilon_{est} \quad \text{(by lemma \ref{lem:qerror})}\\
   & \leq \underset{\lambda \in R_{+}^{2}, ||\lambda||_1 = B}{\max} L(\tilde{q},\lambda) - \underset{q \in \Delta(\M)}{\min} L(q,\tilde{\lambda}) + \epsilon_{est}\\
   &\leq \epsilon_{ol} + \epsilon_{est} \quad \text{(by lemma \ref{lem:primdualgap}).}
\end{align*}
\begin{align*}
\text{Thus, }L(\bar{q},\tilde{\lambda}) &\leq L(q^{f},\tilde{\lambda}) + \epsilon_{ol} + \epsilon_{est} \\
    &= C(q^{f}) + \tilde{\lambda}[1]( D(q^{f})-l) + \epsilon_{ol} + \epsilon_{est} \\
    &\leq C(q^{f}) + \epsilon_{ol} + \epsilon_{est} \quad  (\text{as } D(q^{f}) \leq l ). 
\end{align*}
Putting the above and \eqref{eq:lem3_1} together, we obtain
\begin{align*}
    B(D(\tilde{q})-l) &\leq \tilde{\lambda}[1] (D(\bar{q})-l) +  \epsilon_{ol}/2 + \bar{C}\epsilon_{oe} \\
    &= L(\bar{q},\tilde{\lambda}) - C(\bar{q}) + \epsilon_{ol}/2 + \bar{C}\epsilon_{oe} \\
    &\leq C(q^{f}) + \epsilon_{ol} + \epsilon_{est} - C(\bar{q}) + \epsilon_{ol}/2 + \bar{C}\epsilon_{oe} \\
    &\leq 2(\bar{C}(H+1) + \epsilon_{ol} + \epsilon_{est}).
\end{align*}
We then have $ D(\tilde{q}) \leq l + \frac{ 2(\bar{C}(H+1) + \epsilon_{ol} + \epsilon_{est})}{B} $.

We now show that the the objective $C(\tilde{q})$ returned by $\tilde{q}$ is close to that of $q^{*}$ (occupancy measure associated with optimal policy $\pi^{*}$).
Since $q_t$ is an $\epsilon_{br}$-optimal best response with respect to $\lambda_t$, we have 
\begin{align*}
  C(q_t)+\lambda_{t1}(D(q_t) - l) &\leq C(q^{*})+\lambda_{t1}(D(q^{*}) - l) +\epsilon_{br} \\
  &\leq C(q^{*}) + \epsilon_{br} \quad \text{($q^{*}$ is feasible).}
 \end{align*}
We then conclude that $\frac{1}{T}\sum_{t}L(q_{t},\lambda_{t}) \leq C(q^{*}) + \epsilon_{br} $ and by Lemma~\ref{lem:qerror}, this implies that $\frac{1}{T}\sum_{t}L(\hat{q}_{t},\lambda_{t}) \leq C(q^{*}) + \epsilon_{br} +\epsilon_{est} $ holds. Further, by the no-regret property of the EG algorithm, we obtain: 
$\frac{1}{T}\sum_{t}L(\hat{q}_{t},\lambda_{t})$
\begin{align*}
     &\geq  \underset{\lambda \in R_{+}^{2}, ||\lambda||_1 = B}{\max} L(\tilde{q},\lambda) - \frac{o(T)}{T}\\
    &= \underset{\lambda \in R_{+}^{2}, ||\lambda||_1 = B}{\max} \left[ C(\tilde{q}) + \lambda[1] (D(\tilde{q})-l)\right]- \frac{o(T)}{T}\\
    &\geq C(\tilde{q})- \epsilon_{reg} \\
    &\text{ (by setting $\lambda[1] = 0$ if $D(\tilde{q}) < l$ and $B$ otherwise)}.
\end{align*}
Finally, $ C(\tilde{q})- \epsilon_{reg} \leq \frac{1}{T}\sum_{t}L(\hat{q}_{t},\lambda_{t}) \leq C(q^{*}) + \epsilon_{br} +\epsilon_{est}$ holds, which  implies $C(\tilde{q}) \leq C(q^{*}) + \epsilon_{reg} + \epsilon_{br} +\epsilon_{est} $.

\subsection{Proof of Lemma \ref{lem:occuerror}}

Let $\epsilon_{hsa} = \left|\bar{q}_{h}(s,a) - \tilde{q}_{h}(s,a)\right|$, $\epsilon_{hs} = \sum_{a \in \A} \epsilon_{hsa} $, and $\epsilon_{h} = \sum_{s \in \state} \epsilon_{hs}$. Since $\|\bar{q} - \tilde{q}  \|_{1} \leq \epsilon_{oe}$, we have $\sum_{h}\epsilon_{h} \leq \epsilon_{oe} $. \\\\
For $h=0$, we also have $\sum_{s\in \state}\sum_{a \in \A}|\bar{q}_{0}(s,a) - \tilde{\tilde{q}}_{0}(s,a)| $
\allowdisplaybreaks\begin{align*}
    &= \
    \sum_{s\in \state}\sum_{a \in \A} \mathds{1}(s = s_0) |\bar{q}_{0}(s,a) - \tilde{{q}}_{0}(s,a)|  \\
   &\text{(by construction of $\tilde{\pi}$ and definition of $\tilde{\tilde{q}}_{0}(s,a)$)} \\
   &=\sum_{a \in \A} |\bar{q}_{0}(s_0,a) - \tilde{q}_{0}(s_0,a)|  \\
   &\text{(as $\bar{q}_{0}(s,a), \tilde{{q}}_{0}(s,a) = 0$ for $s\neq s_0$)}\\
    &\leq \epsilon_{0} \leq 2\epsilon_{0}.
\end{align*}

Let $e_{hsa} = |\bar{q}_{h}(s,a) - \tilde{\tilde{q}}_{h}(s,a)|$ and let us assume that, for $h = k, 0 \leq k < H $,
$\sum_{s\in \state}\sum_{a \in \A} e_{ksa} \leq 2\sum_{i=0}^{k}\epsilon_{i}$.\\

Then, for $h=k+1$, we obtain $|\bar{q}_{k+1}(s,a) - \tilde{\tilde{q}}_{k+1}(s,a)|= $
\allowdisplaybreaks\begin{align*}
     & =|\bar{q}_{k+1}(s,a) - \tilde{\pi}_{k+1}(a|s)\tilde{\tilde{q}}_{k+1}(s)|\\
    &= |\bar{q}_{k+1}(s,a) - \tilde{\pi}_{k+1}(a|s)\bar{q}_{k+1}(s) + \\ 
    & \tilde{\pi}_{k+1}(a|s)\bar{q}_{k+1}(s) -  \tilde{\pi}_{k+1}(a|s)\tilde{\tilde{q}}_{k+1}(s)| \\
    &= |\bar{q}_{k+1}(s,a) - \frac{\tilde{q}_{k+1}(s,a)}{\tilde{q}_{k+1}(s)}\bar{q}_{k+1}(s) + \\
    & \tilde{\pi}_{k+1}(a|s) (\bar{q}_{k+1}(s) - \tilde{\tilde{q}}_{k+1}(s))| \text{ (by construction of $\tilde{\pi}$)}\\
    &\leq |\bar{q}_{k+1}(s,a) - \frac{\tilde{q}_{k+1}(s,a)}{\tilde{q}_{k+1}(s)}\bar{q}_{k+1}(s)| + \tilde{\pi}_{k+1}(a|s) |\bar{q}_{k+1}(s) - \tilde{\tilde{q}}_{k+1}(s)|\\
    &=|\bar{q}_{k+1}(s,a) - \frac{\tilde{q}_{k+1}(s,a)}{\tilde{q}_{k+1}(s)}(\tilde{q}_{k+1}(s)-\tilde{q}_{k+1}(s) +\bar{q}_{k+1}(s))| + \\ &+ \tilde{\pi}_{k+1}(a|s) |\bar{q}_{k+1}(s) - \tilde{\tilde{q}}_{k+1}(s)|\\
    &\leq |\bar{q}_{k+1}(s,a) - \tilde{q}_{k+1}(s,a)| + \frac{\tilde{q}_{k+1}(s,a)}{\tilde{q}_{k+1}(s)} |\bar{q}_{k+1}(s) - \tilde{q}_{k+1}(s) | + \\ &+ \tilde{\pi}_{k+1}(a|s) |\bar{q}_{k+1}(s) - \tilde{\tilde{q}}_{k+1}(s)|.
\end{align*}

Summing over $a \in \A$, we get
\allowdisplaybreaks\begin{align*}
&\sum_{a\in \A} |\bar{q}_{k+1}(s,a) - \tilde{\tilde{q}}_{k+1}(s,a)| \\
&\leq \sum_{a\in \A}\epsilon_{(k+1)sa} + \frac{\sum_{a\in \A} \tilde{q}_{k+1}(s,a)}{\tilde{q}_{k+1}(s)} |\bar{q}_{k+1}(s) - \tilde{q}_{k+1}(s) | + \\ &+ \sum_{a\in \A} \tilde{\pi}_{k+1}(a|s) |\bar{q}_{k+1}(s) - \tilde{\tilde{q}}_{k+1}(s)| \\
&= \sum_{a\in \A}\epsilon_{(k+1)sa} + |\bar{q}_{k+1}(s) - \tilde{q}_{k+1}(s) | + \\ 
&+|\bar{q}_{k+1}(s) - \tilde{\tilde{q}}_{k+1}(s)| \sum_{a\in \A} \tilde{\pi}_{k+1}(a|s)\\
&\leq \sum_{a\in \A}\epsilon_{(k+1)sa} + \sum_{a\in \A}\epsilon_{(k+1)sa} + |\bar{q}_{k+1}(s) - \tilde{\tilde{q}}_{k+1}(s)| \\
&\text{(by triangular inequality)}\\
&= 2\epsilon_{(k+1)s} + |\bar{q}_{k+1}(s) - \tilde{\tilde{q}}_{k+1}(s)|\\
&= 2\epsilon_{(k+1)s} + |\sum_{s'\in \state,a' \in \A} p_{k}(s|s',a')(\bar{q}_{k}(s',a') - \tilde{\tilde{q}}_{k}(s',a'))|\\
&\leq 2\epsilon_{(k+1)s} + \sum_{s'\in \state,a' \in \A} p_{k}(s|s',a') |\bar{q}_{k}(s',a') - \tilde{\tilde{q}}_{k}(s',a')|.
\end{align*}

Summing over $s \in \state$, we get $\sum_{s \in \state} \sum_{a\in \A} |\bar{q}_{k+1}(s,a) - \tilde{\tilde{q}}_{k+1}(s,a)|$
\allowdisplaybreaks\begin{align*}
    &\leq 2 \sum_{s \in \state}\epsilon_{(k+1)s} +   \sum_{s \in \state} \sum_{s'\in \state,a' \in \A} p_{k}(s|s',a') |\bar{q}_{k}(s',a') - \tilde{\tilde{q}}_{k}(s',a')|\\
    &\text{(by flow conservation property of occupancy measure)}\\
    &= 2\epsilon_{k+1} +  \sum_{s'\in \state,a' \in \A} |\bar{q}_{k}(s',a') - \tilde{\tilde{q}}_{k}(s',a')|\sum_{s \in \state}p_{k}(s|s',a')\\
    &= 2\epsilon_{k+1} +  \sum_{s'\in \state,a' \in \A} e_{ks'a'}\\
    &\leq 2\epsilon_{k+1} + 2\sum_{i=0}^{k}\epsilon_{i} \\
    &= 2\sum_{i=0}^{k+1}\epsilon_{i}.
\end{align*}

Thus, by induction, we have that for $0 \leq h \leq H$,
$ \sum_{s\in \state}\sum_{a \in \A} |\bar{q}_{h}(s,a) - \tilde{\tilde{q}}_{h}(s,a)| \leq 2\sum_{i=0}^{h}\epsilon_{i} $ holds. This implies
$\|\bar{q} - \tilde{\tilde{q}}  \|_{1} = \sum_{h=0}^{H} \sum_{s\in \state}\sum_{a \in \A} |\bar{q}_{h}(s,a) - \tilde{\tilde{q}}_{h}(s,a)| \leq 2\sum_{h=0}^{H}\sum_{i=0}^{h}\epsilon_{i}
    \leq 2(H+1)\epsilon_{oe} $.

\end{document}